\documentclass[journal=jacsat,manuscript=article]{achemso}

\usepackage{chemformula} 
\usepackage[T1]{fontenc} 
\usepackage{siunitx} 
\newcommand{\code}[1]{\colorbox{gray!10}{\texttt{#1}}} 
\usepackage[colorlinks=true, linkcolor=blue, citecolor=blue, urlcolor=blue]{hyperref} 
\usepackage{placeins} 

\author{T. J. Sturges}
\affiliation{Institute of Theoretical Solid State Physics, Karlsruhe Institute of Technology (KIT), 76131 Karlsruhe, Germany}
\email{thomas.sturges@kit.edu}

\author{M. Nyman}
\affiliation{Institute of Nanotechnology, Karlsruhe Institute of Technology (KIT), 76131 Karlsruhe, Germany}

\author{S. Kalt}
\affiliation{Institute of Applied Physics (APH), Karlsruhe Institute of Technology (KIT), 76131 Karlsruhe, Germany}

\author{K. P\"{a}lsi}
\affiliation{Department of Applied Physics, Aalto University, PO Box 13500, FI-00076 Aalto, Finland}

\author{P. Hilden}
\affiliation{Department of Applied Physics, Aalto University, PO Box 13500, FI-00076 Aalto, Finland}

\author{M. Wegener}
\affiliation{Institute of Nanotechnology, Karlsruhe Institute of Technology (KIT), 76131 Karlsruhe, Germany}
\alsoaffiliation{Institute of Applied Physics (APH), Karlsruhe Institute of Technology (KIT), 76131 Karlsruhe, Germany}

\author{C. Rockstuhl}
\affiliation{Institute of Theoretical Solid State Physics, Karlsruhe Institute of Technology (KIT), 76131 Karlsruhe, Germany}
\alsoaffiliation{Institute of Nanotechnology, Karlsruhe Institute of Technology (KIT), 76131 Karlsruhe, Germany}

\author{A. Shevchenko}
\affiliation{Department of Applied Physics, Aalto University, PO Box 13500, FI-00076 Aalto, Finland}
\email{andriy.shevchenko@aalto.fi}

\title{Inverse-designed 3D laser nanoprinted phase masks to extend the depth of field of imaging systems}

\abbreviations{PSF}
\keywords{optical imaging, depth of field, phase mask, inverse design, 3D laser nanoprinting}

\begin{document}

\begin{tocentry}

\centering
\vspace{5pt}
\includegraphics[]{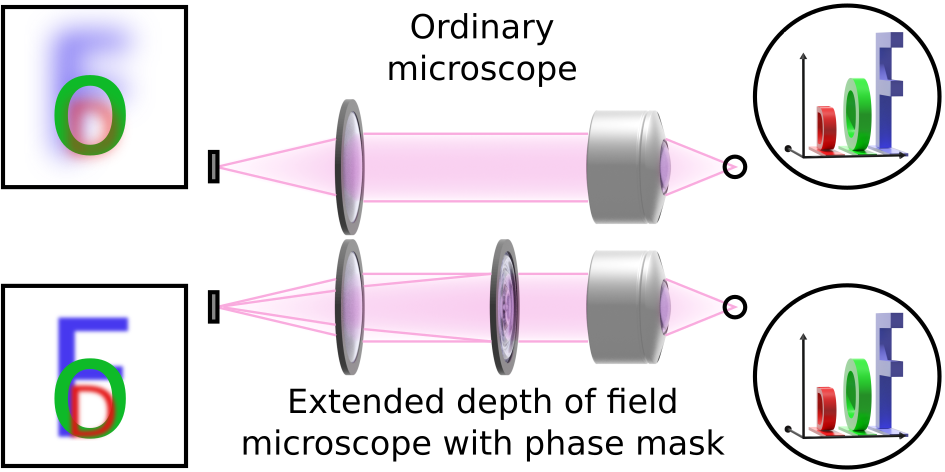}

\end{tocentry}

\begin{abstract}
\noindent In optical imaging, achieving high resolution often comes at the expense of a shallow depth of field. This means that when using a standard microscope, any minor movement of the object along the optical axis can cause the image to become blurry. To address this issue, we exploit inverse design techniques to optimise a phase mask which, when inserted into a standard microscope, extends the depth of field by a factor of approximately four without compromising the microscope's resolution. Differentiable Fourier optics simulations allow us to rapidly iterate towards an optimised design in a hybrid fashion, starting with gradient-free Bayesian optimisation and proceeding to a local gradient-based optimisation. To fabricate the device, a commercial two-photon 3D laser nanoprinter is used, in combination with a two-step pre-compensation routine, providing high fabrication speed and much better than subwavelength accuracy. We find excellent agreement between our numerical predictions and the measurements upon integrating the phase mask into a microscope and optically characterising selected samples. The phase mask enables us to conduct simultaneous multiplane imaging of objects separated by distances that cannot be achieved with the original microscope.
\end{abstract}


\section{Introduction}

A frequent limitation of high-resolution optical imaging systems is their shallow depth of field. As a consequence, even a minor shift of an object along the optical axis results in a blurred image. Multiple techniques have been developed to extend the depth of field. These generally involve spatial modulation of the light in the system; either of the amplitude \cite{Hilden2023, Hegedus1985, OjedaCastaneda1986, OjedaCastaneda1987}, or the phase profile  \cite{BenEliezer2003, Davidson1991, Zhao2024, Kolodziejczyk1990, Kai2023, Bayati2020, Bayati2022, Seong2023}. For example, in a previous work \cite{Hilden2023}, some of the authors herein used a thin ring-shaped slit (an annular aperture placed in the Fourier plane of the system) leading to a point spread function (PSF) of nearly non-diverging Bessel-like beams \cite{McGloin2005}. In this case, the image of an arbitrary object is rather insensitive to the position of the detectors; and by reciprocity, if the object is shifted along the optical axis, its image at the image plane remains sharp over greater displacements. However, this and other approaches that use spatial filtering \cite{Hegedus1985, OjedaCastaneda1986, OjedaCastaneda1987} have some drawbacks. Firstly, Bessel beams have pronounced side-lobes, which results in a reduced image contrast and artefacts such as halos. Secondly, an annular or other similar aperture results in low transmission, which brings problems such as increased noise and the need for longer exposure times. For example, if the object is highly transparent, the field in the Fourier plane is well localised on the optical axis and thus is essentially completely blocked by the aperture.

To circumvent the problem of low transmission, a phase mask can be used instead. Some methods are based on all-optical phase modulation \cite{BenEliezer2003, Kolodziejczyk1990, Davidson1991, Zhao2024, Kai2023, Bayati2020, Bayati2022}, whereas others use an optical-digital method that relies on cross-calibrated optical and digital signal processing \cite{Dowski1995, Bradburn1997, Sherif2004, Chu2008}. These methods include incoherent digital holography, which significantly extends the effective depth of field \cite{Nobukawa2019, Rai2021}. The use of neural networks has also been explored, both for upscaling the resolution \cite{Rivenson2017}, as well as jointly optimising the optical and digital processing stages \cite{Akpinar2021, Seong2023}.

A fundamental question is how one should design a phase mask to provide the best improvement in the depth of field, without compromising the resolution. In the last decade or so, photonic inverse design has been used to create many different optical structures with tailored functionalities \cite{Molesky2018, Augenstein2020, Minkov2020, Ahn2022, Forestiere2016, Keawmuang2024, So2020, Wiecha2021, Wiecha2022, Bayati2020, Bayati2022}. This often exploits fully differentiable software packages with built-in adjoint solvers that enable one to optimise devices with a very large number of design variables, allowing practically arbitrary design freedom. On the other hand, global optimisation techniques \cite{Schneider2019} can sometimes offer more optimal solutions, but are computationally impractical for design spaces with more than just a handful of variables. In this work, we used a hybrid global-then-local inverse design method to design a phase mask that extends the depth of field of a microscope that uses spatially incoherent illumination. We find this inverse design method convenient, not only because of its speed and quality of the results, but also because it allows the designer to make tradeoffs and adjust the balance between, for instance, achieving a larger depth of field enhancement at the expense of image quality. 

The designed mask was subsequently fabricated with a 3D laser nanoprinter \cite{multi-focus, Farsari2009, Zyla2024}, and optical measurements were conducted to verify its performance. We found that the depth of field was extended by a factor of approximately 4, in excellent agreement with the design and simulations. As our design method yields slowly-varying phase profiles, 3D nanoprinting allows the fabrication of large-area freeform structures that will be essential when extending the depth of field of demanding optical systems, such as those with large numerical apertures. We note that whilst we chose to encode our phase mask in the height profile of a glass plate, this could be alternatively implemented in other physical platforms, such as metamaterials and metasurfaces that can also implement arbitrary phase transparencies (see, e.g., Ref.~\cite{overvigDielectricMetasurfacesComplete2019}). Indeed, in a similar vein, inverse techniques have been used to extend the depth of \textit{focus} of metalenses \cite{Bayati2020, Bayati2022}. In such works, it is the lens of the optical system that is nano- or microstructured, whereas in our approach the phase mask is a separate optical element. In contrast to the metalens approach, which usually relies on electron-beam or deep ultraviolet mask lithography, 3D laser printing of phase masks allows for a rapid iteration of designs. Hence, imperfections in the optical device can be compensated for and theoretically driven design changes incorporated in the fabrication process with comparably low effort. In addition, the combination of a slowly-varying phase profile and the possibility to 3D laser print optical-grade surfaces with low surface roughness significantly improves the power transmission as losses due to scattering or higher-order diffraction are minimised.

\section{Results}

\subsection{Overview of the setup}
\FloatBarrier

\begin{figure}[hbt!]
\centering 
\includegraphics[width=\textwidth]{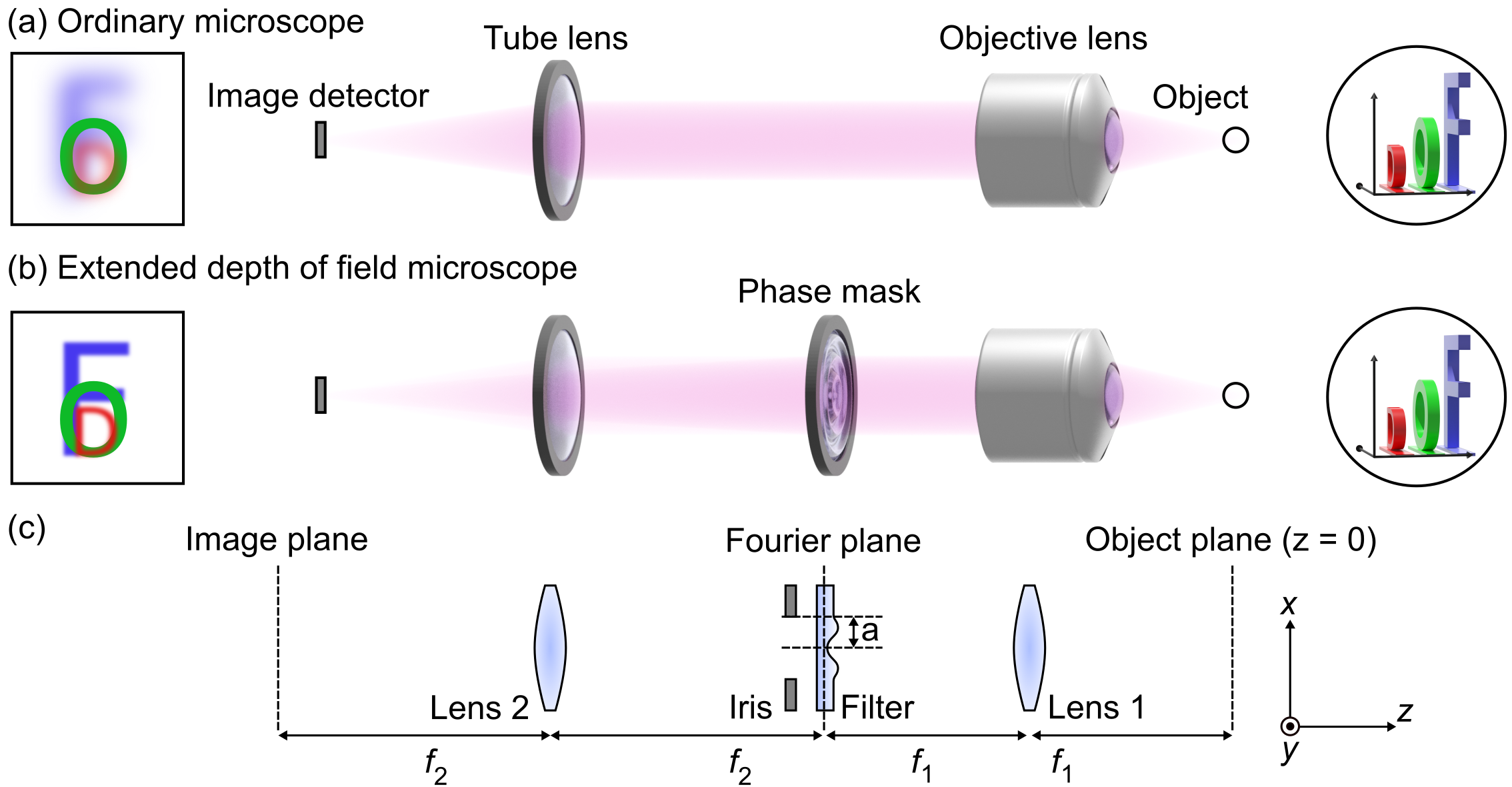}
\caption{A conceptual schematic of the imaging system (a) without and (b) with the optimised phase mask. We imagine an `object' made of 3D letters at different distances from the object plane, with the ``O" brought into focus. In (a) we see that the letters ``D" and ``F" are blurry, whereas in (b) the phase mask provides an extended depth of field, enabling these letters to be resolved. Further details are shown in (c); here, lenses 1 and 2 are the objective and tube lenses, respectively, with focal lengths $f_1$ and $f_2$. Note that an iris is placed immediately after the phase mask.} 
\label{fig:setup}
\end{figure}

Our goal is to extend the depth of field of an incoherent imaging system using a phase mask. We consider a 4$f$-type imaging system, as it is the simplest and most efficient system that allows for spatial phase modulation. In Fig.~\ref{fig:setup}, we show a schematic of the setup and a conceptual overview of our research goal.
A plain 4$f$ system (Fig.~\ref{fig:setup}a) consists of an objective lens and a tube lens separated by a distance $f_1 + f_2$, where $f_1 = \qty{45}{\milli\meter}$ (resp. $f_2 = \qty{200}{\milli\meter}$) is the focal length of the objective (resp. tube) lens. Hence, the two lenses share a common focal plane, referred to as the Fourier plane. By placing the detector at the back focal plane of the tube lens (the image plane), and the object at the front focal plane of the objective lens (the object plane), we obtain an optical setup that behaves as an ordinary microscope. The magnification factor of the microscope is $M = 4.44$. We denote the optical axis as the $z$-axis, with the origin at the object plane, and the positive direction away from the tube lens.

\FloatBarrier

To emphasise our research goal, we consider a specimen consisting of three objects (shown as 3D letters in Fig.~\ref{fig:setup}) at different planes along the optical path, with the central object (the letter ``O'') placed at the principal object plane, and thus in focus. The other two objects (the letters ``D'' and ``F'') will be out of focus if their distance is greater than the depth of field of the microscope. Therefore, if we would like to simultaneously resolve these objects, we must increase the depth of field of the imaging system.

\FloatBarrier
\subsection{Simulations and mask design}
\label{results:simulations}
\FloatBarrier

We use inverse design techniques to optimise the phase profile of a radially symmetric phase mask placed in the Fourier plane to provide an extended depth of field, as illustrated in Fig.~\ref{fig:setup}b. The phase mask consists of a free-form surface structure made of two-photon-polymerised IP-S photoresist on a glass substrate. It has a refractive index of $n=1.509$ \cite{Schmid2019}, and a radius of $a=\qty{1}{\milli\meter}$. An iris with the same radius is placed just after it, at a distance of \qty{2}{\milli\metre}, to prevent light from passing through the unstructured part and contributing to the image. To control the phase profile, we optimise the height profile of the free-form surface. We optimise the device for a single wavelength of $\lambda = \qty{633}{\nano\meter}$; however, we find that the imaging performance remains consistent across different wavelengths within the visible range. (see Supporting Information for details). We simulate the imaging system using Fourier optics, and model the optical elements within the thin element approximation (see Methods section). Our code is written using JAX \cite{jax2018github}, a software package that can automatically differentiate native Python and Numpy code. This enables us to obtain the derivatives of simulation results with respect to the variables that parameterise the design of the phase mask. We obtain all derivatives simultaneously from a single ``backwards'' pass through the simulation code, with a computational complexity that is comparable with the original function, regardless of the number of design variables \cite{Griewank2008, Baydin2018}. Therefore, we can specify a figure of merit (objective function) and then use gradient-based optimisation to iterate towards an optimised design. However, the final solution is influenced by the initial phase profile provided to the gradient-based optimiser, which can get stuck in suboptimal local optima. Therefore, we complement this approach with an initial global optimisation that uses fewer design variables, in order to find a good starting design for the gradient-based optimisation. Specifically, our three-step approach is as follows: first, we parameterise the surface topography of the (radially symmetric) phase mask using a 6th-order Chebyshev polynomial and use Bayesian optimisation to obtain an initial design; then we refine the topography using gradient descent with a 10th order Chebyshev polynomial; finally, we perform gradient descent to optimise the surface ring-by-ring using an equispaced radial grid with 50 points. 

\begin{figure}[ht!]
\centering 
\includegraphics[width=\textwidth]{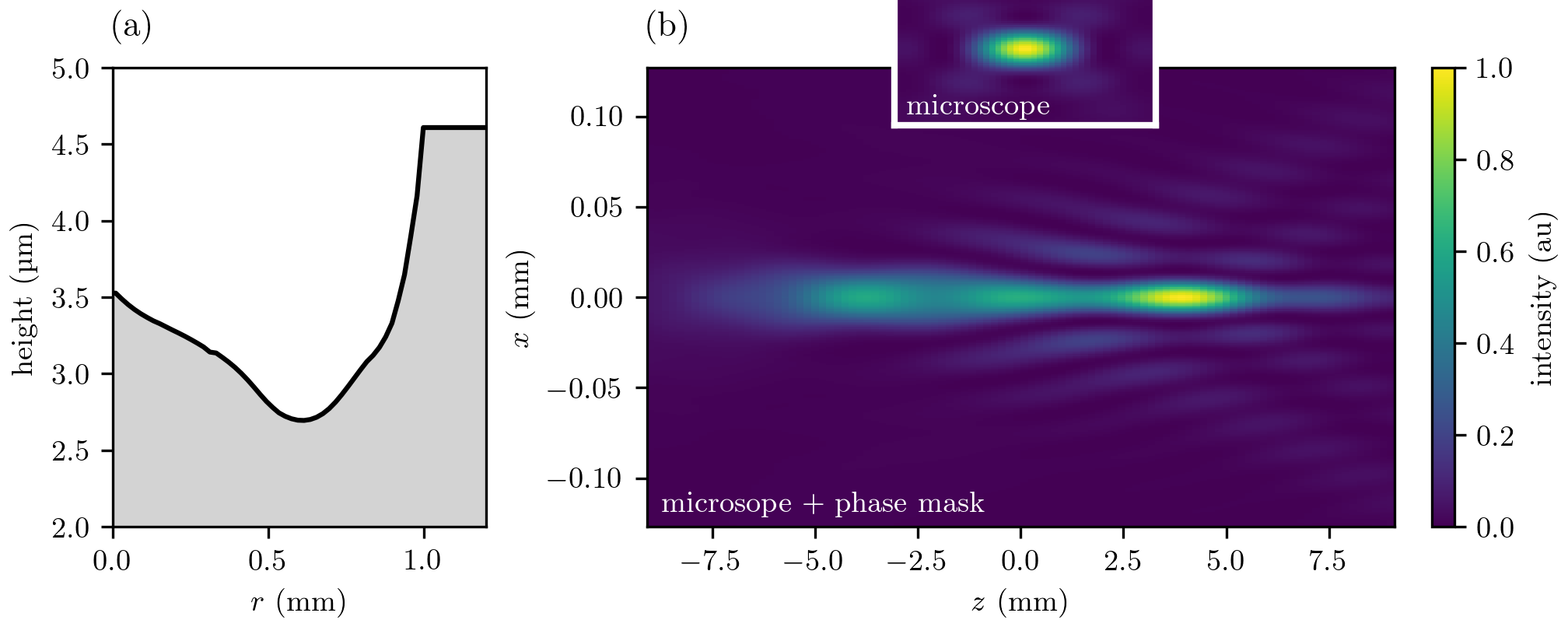}
\includegraphics[width=\textwidth]{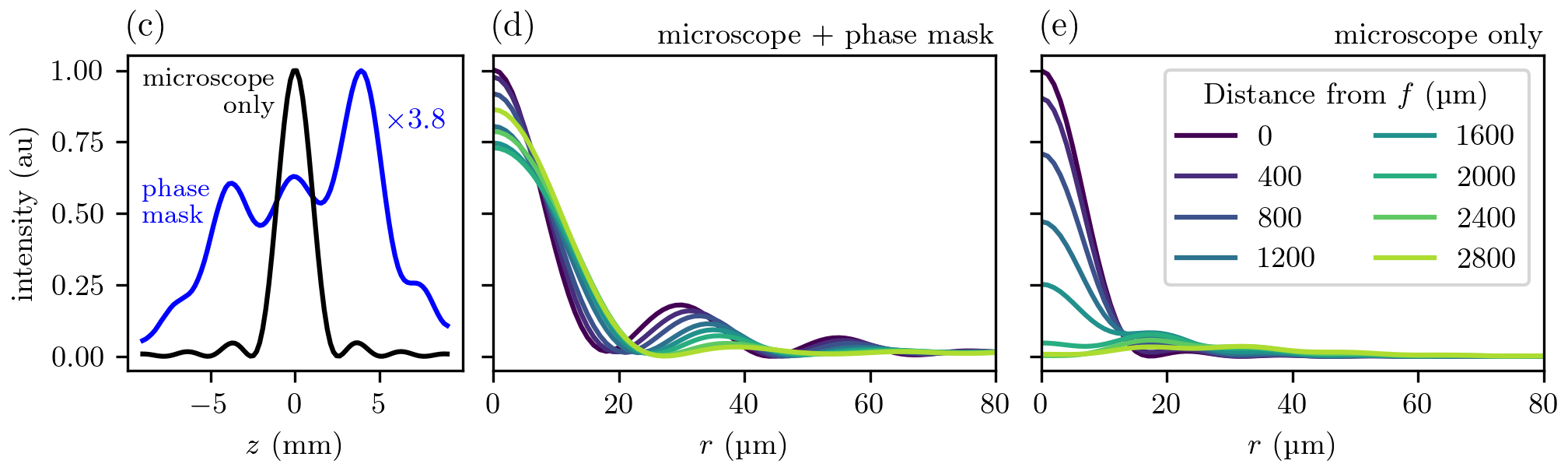}
\caption{The optimised phase mask design and simulations showing an increased depth of field. (a) The optimised height profile of the phase mask as a function of the radial coordinate. For $r > \qty{1}{\milli\meter}$ the height is constant. (b) Simulated point spread function (PSF) on the object side of the imaging setup, exhibiting an extended depth of field; $z$ is the distance from the focal plane along the optical axis. The inset shows the PSF of the imaging system without the phase mask (the image is shifted vertically for clarity, but the size scale is not changed). (c) Cuts of the PSF along the optical axis for the imaging system without (black) and with (blue) the phase mask. (d-e) Cuts of the PSF along the radial axis for (d) the optimised design, and (e) the microscope only. These figures show that the phase mask increases the depth of field at the cost of reduced peak intensity, more pronounced side lobes, and increased axial inhomogeneities.} 
\label{fig:simulation}
\end{figure}

In our simulations, we consider the propagation of light from a point source located at the image plane, backwards through the imaging system, and record the field intensities in the vicinity of the object plane. This might seem counterintuitive, but due to optical reciprocity, this approach allows us to encode the target depth of field directly into the PSF of this reversed imaging system (see Supporting Information for further explanation). After initial experimentation, we settled on an objective function that consists of a weighted sum of two sub-objectives with opposite signs. The first sub-objective is the sum of the field intensities at equidistant points on a `thick' line, that runs along the optical axis, centered at the focal point, with a length equal to the target depth of field. This rewards the optimiser for making a long PSF with high resolution. The second sub-objective is the sum of the absolute differences along the same line. The second objective has a negative weight, thus penalising intensity variations along the line, and helping to make the PSF smooth. By `thick', we specifically mean that at each point along the optical axis, we multiply the corresponding transverse field intensity profile with a narrow Gaussian centered at that point, thus avoiding numerical issues arising from discretisation.

In Fig.~\ref{fig:simulation} we show the optimised design (Fig.~\ref{fig:simulation}a) and the results of our field simulations. In Figs.~\ref{fig:simulation}b and \ref{fig:simulation}c, we can see how the inclusion of the phase mask acts to elongate the PSF of the system, signalling an increased depth of field. However, we also observe an inhomogeneous intensity profile along the optical axis, and of course, a reduced overall intensity due to the redistribution of optical power across a wider depth of field. Note also that the intensity and location of the side lobes appear to depend on the position along the optical axis. Figures~\ref{fig:simulation}d-e highlight how the phase mask results in a central peak that remains significant for larger displacements, at the cost of more pronounced side lobes. To quantify the depth of field, we define it as the distance between the two optical planes at which the standard deviation of the PSF (calculated in the $xy$-plane) is twice its value at the object plane. There is one such plane in both the positive and negative $z$-direction. In other words, the depth of field is the region within which the PSF remains less than double its smallest value. We determine the improvement factor in the depth of field as approximately 4.3. At the focal point, the full-width half-maximum (FWHM) of the imaging system is increased by 13\% by the inclusion of the phase mask. In this sense the resolution is minimally impacted (see Supplementary Information for further analysis of the FWHM). We emphasise that for a microscope using spatially incoherent illumination, the side lobes are less important than for a coherent system, because the images are formed by displaced PSFs that do not interfere \cite{Hilden2023}.

\FloatBarrier
\subsection{Fabrication and characterisation of the phase mask}
\FloatBarrier

In order for the phase mask to perform as designed, its height profile should match the calculated surface profile as closely as possible. Specifically, only deviations much smaller than the considered wavelengths are acceptable. To achieve the required surface quality, the phase mask is fabricated using two-photon grayscale lithography \cite{multi-focus, Farsari2009, Zyla2024} (see the Methods section for details). This technique is a laser-based 3D fabrication technique that uses short laser pulses to induce polymerisation of a photosensitive resin. 

\begin{figure}[ht!]
\centering 
\includegraphics[width=\textwidth]{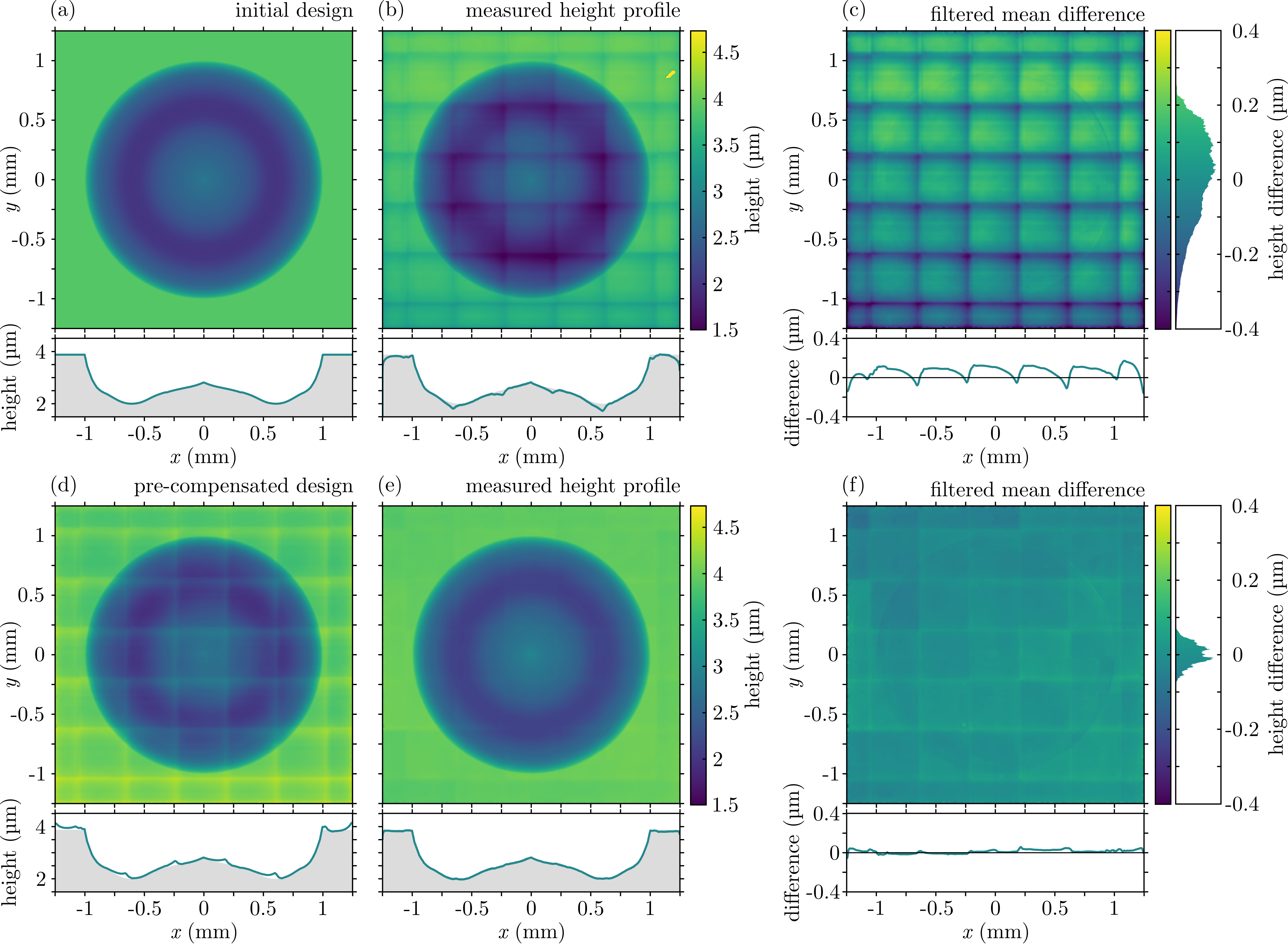}
\caption{Demonstration of the effectiveness of the two-photon lithography with pre-compensation routine, used to fabricate the target phase mask with high accuracy. The first row of panels displays the results obtained with the initial design. The second row shows how the shape accuracy of the fabricated structures improves when using a pre-compensated target design. (a) The initial design used to print the first iteration of the device. (b) Microscopy measurements of the height profile of a printed device. (c) To reduce the influence of random defects and measurement errors, we calculate the filtered mean difference between the initial design and an ensemble of four independently fabricated devices. (d) The second-iteration design with pre-compensation for stitching and other systematic structure imperfections. (e) The height profile of a second-iteration design. (f) The filtered mean difference for the ensemble of second-iteration devices. We see that the mean differences are all much smaller than visible wavelengths, which is important for the device to perform as simulated. In all figures, we show a cut along the diameter of the device. All shown measurements are offset-, tilt-, and rotation-corrected.} 
\label{fig:fabrication}
\end{figure}

To account for several systematic and statistical imperfections, a two-step printing process is used. In the first step, we naively print the device using the optimised design without modification. The top row of Fig.~\ref{fig:fabrication} compares the measured height profile between the fabricated device (Fig.~\ref{fig:fabrication}b) and the target design (Fig.~\ref{fig:fabrication}a). We observe a square grid of stitching artefacts due to the limited size of the field of view of the printing setup. To reduce this and other systematic imperfections, a pre-compensation routine is used to print a next iteration device. This routine, available in an open-source software tool \cite{pre-compensation}, calculates the difference map between the measured height profile and the target structure. This difference map is then used to calculate a modified target design that pre-compensates for the systematic imperfections introduced during printing. However, statistical imperfections cannot be eliminated by this routine due to their random nature. Hence, it is important to ensure that their influence is not incorporated in the new pre-compensated design. For this reason, four phase masks with identical input designs and printing parameters are printed and independently characterised to calculate a mean difference map. Remaining local deviations with strong height gradients in the mean difference map, caused e.g. by dust, are removed with a Gaussian filter so that they are not propagated to the new design. Such a filtered difference map is shown in Fig.~\ref{fig:fabrication}c.

In Fig.~\ref{fig:fabrication}d, we show the second iteration target design, with the effect of the stitching artefacts clearly visible in this pre-compensated design. The second iteration device, shown in Fig.~\ref{fig:fabrication}e, is a much better fit to the original design, as can be seen by examining the filtered mean difference map in Fig.~\ref{fig:fabrication}f (also calculated with four independently fabricated devices). In the first iteration devices, the average absolute difference is $\mu_\text{abs} = \qty{101.1}{\nano\meter}$, with a standard deviation of $\sigma = \qty{125.0}{\nano\meter}$ and 99\% of the mean differences below \qty{334}{\nano\meter}; thus, the differences are mostly smaller than visible wavelengths, but a non-negligible fraction of them are of comparable magnitude. In the second iteration devices, we obtain $\mu_\text{abs} = \qty{23.8}{\nano\meter}$, $\sigma = \qty{30.7}{\nano\meter}$, and 99\% of the mean differences below \qty{88}{\nano\meter}; thus, the overwhelming majority of differences are far smaller than visible wavelengths, meaning we can expect the device to perform as it was designed in the simulations (see the Supporting Information for a comparison between simulations with the original design and simulations with the topography of the measured second iteration device).

\FloatBarrier
\subsection{Optical characterisation of the imaging system}
\FloatBarrier

\begin{figure}[ht!]
\centering 
\includegraphics[width=\textwidth]{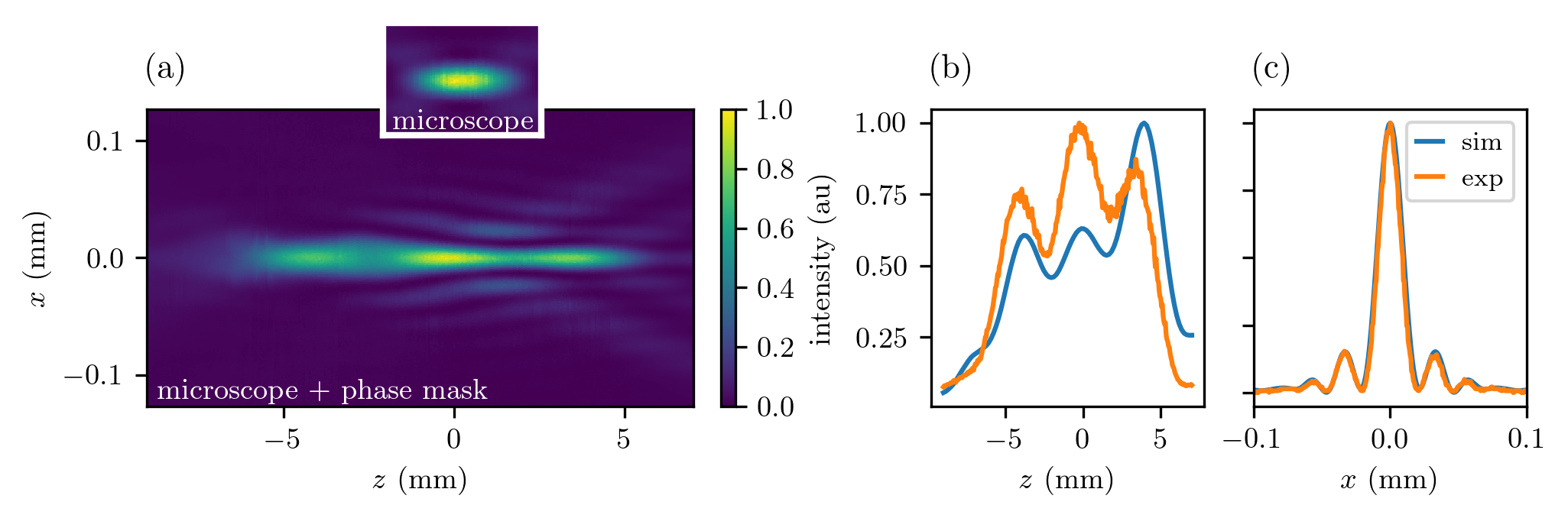}
\caption{Optical characterisation of the device. (a) An image composed of multiple images of a pin-hole aperture displaced away from the object plane, along the optical axis. This approximately corresponds to the point spread function of the imaging system. (b-c) Cuts of the intensity distribution (b) along the optical axis at $r=0$, and (c) along the radial axis at $z=0$. The experimental and calculated curves are represented by the orange and blue lines, respectively. We see good agreement between the measured intensity distribution and the simulated point spread function.} 
\label{fig:verification}
\end{figure}

To verify that the fabricated phase mask provides the extended depth of field, as predicted by simulations, we perform optical measurements of the PSF of the imaging system. We place a \qty{5}{\micro\meter} diameter pinhole aperture (initially) at the image plane. Then, multiple images of the pinhole are taken with an LED (Thorlabs M625L3), at \qty{50}{\micro\meter} steps in the negative and positive $z$-direction (\qty{100}{\micro\meter} steps for the imaging system without the phase mask). The intensity profile in Fig.~\ref{fig:verification}a is a composition of slices of these images. We see that the measured PSF is in strong qualitative agreement with the simulated one (shown in Fig.~\ref{fig:simulation}b), and thus the same conclusions as before apply here. Namely, compared to the ordinary microscope (inset in Fig.~\ref{fig:verification}a), we observe an increased depth of field, as evidenced by the elongated PSF. This comes at the cost of more pronounced side lobes, an inhomogeneous PSF, and a decrease in the intensity at any particular image plane. Figures~\ref{fig:verification}b-c provide a more quantitative comparison between the simulation and measurement of the PSF, where the strong qualitative agreement is further evidenced. From the measurements, we determine the improvement factor in the depth of field as approximately 3.7.

\begin{figure}[ht!]
\centering 
\includegraphics[width=\textwidth]{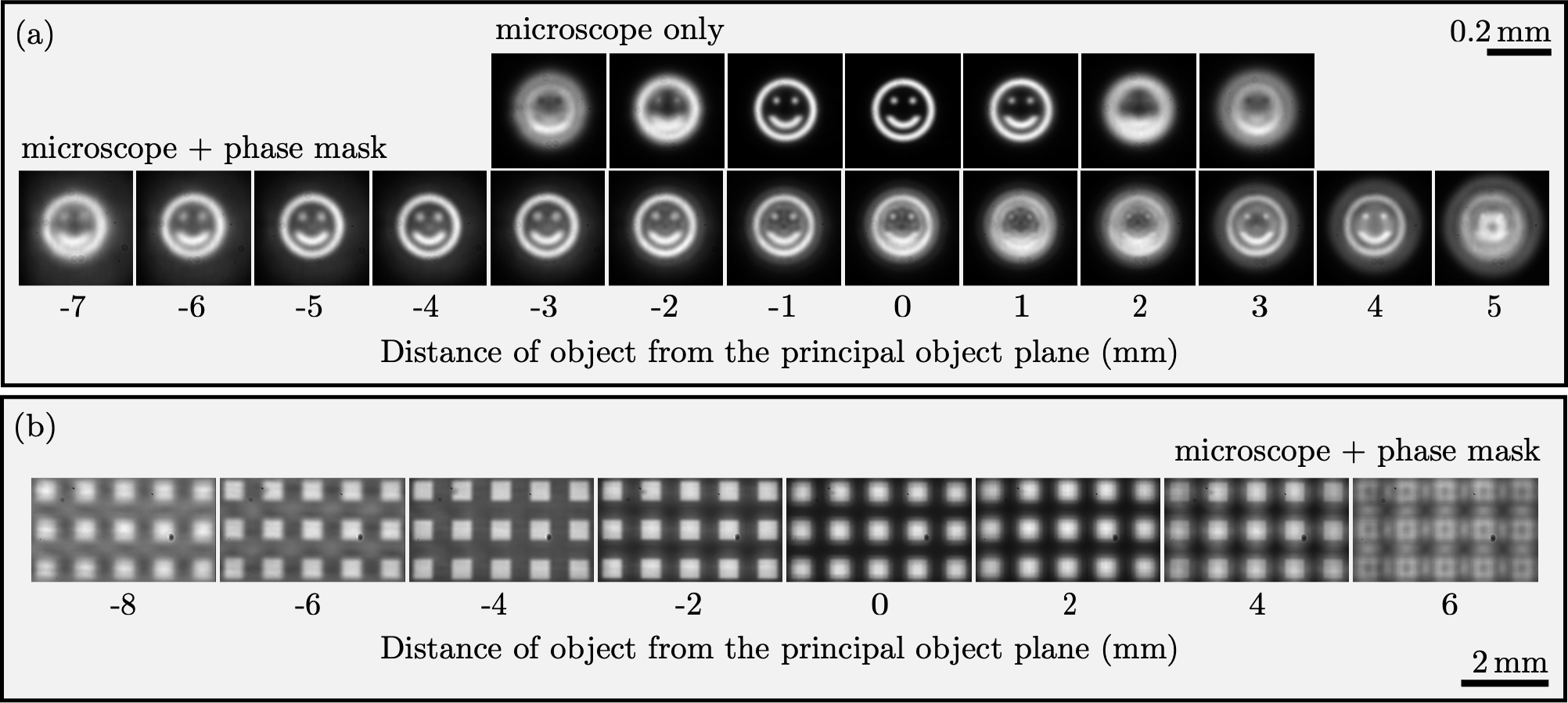}
\caption{Images of transparencies as they are moved out of focus, along the $z$-axis. (a) Images of a `smile' etched into a thin film of aluminium on top of a glass plate. Images are taken at different positive and negative displacements from the principal object plane. With the phase mask inserted (bottom row), we see that the key features remain discernible over larger displacements, as compared to the microscope only (top row). (b) Images of an array of transparent squares. These images are significantly larger than the smile (see scale bars), with the edges far from the optical axis. We see no significant off-axis aberrations at this scale. The images are normalised to their peak intensity.} 
\label{fig:imaging}
\end{figure}

We analyse the `real-world' performance of the imaging system by taking images of example objects at different object planes, with a white-light incoherent source (LEICA CLS 160X), as shown in Fig.~\ref{fig:imaging}. In Fig.~\ref{fig:imaging}a, we compare the images of a `smile' taken with and without the phase mask, at several different displacements away from the principal object plane, with a step size of 1 mm. The smiles are made by etching into a thin aluminium film on top of a glass plate, and have an outer diameter of \qty{200}{\micro\meter}. With the phase mask inserted, we see that the key features remain discernible over larger displacements of the object. However, we observe artefacts such as halos, as well as a reduced contrast. In addition, the imaging performance is asymmetrical with respect to the positive and negative displacement from the focal plane; it appears worse in the positive $z$ direction. In principle, imaging processing algorithms could reduce such artefacts, but this goes beyond the scope of our work. In particular, deconvolution algorithms which make use of the known PSF could be beneficial, but such algorithms should be capable of handling the fact that our PSF is not invariant along the optical axis.

In Fig.~\ref{fig:imaging}b we show images of a `grating', which is made by depositing opaque aluminium stripes on top of a glass plate. The period of the grating is approximately \qty{250}{\micro\meter} and its magnified image fills the entire \qtyproduct{4.22 x 2.38}{\milli\meter} sensor of the camera (Balser ace acA1920-25uc). This larger object serves to show that there are no noticeable off-axis aberrations. Note that a 4$f$ system with the objective lens replaced by an axicon (which produces Bessel beams) would cause large off-axis aberrations \cite{Tanaka2000}, and render the microscope unusable outside of a tiny field of view. Thus, although the PSF of the system appears to be a Bessel-\textit{like} beam structure, our optimisation procedure resulted in a beam structure that is absent of such aberrations.

\begin{figure}[ht!]
\centering 
\includegraphics[width=0.85\textwidth]{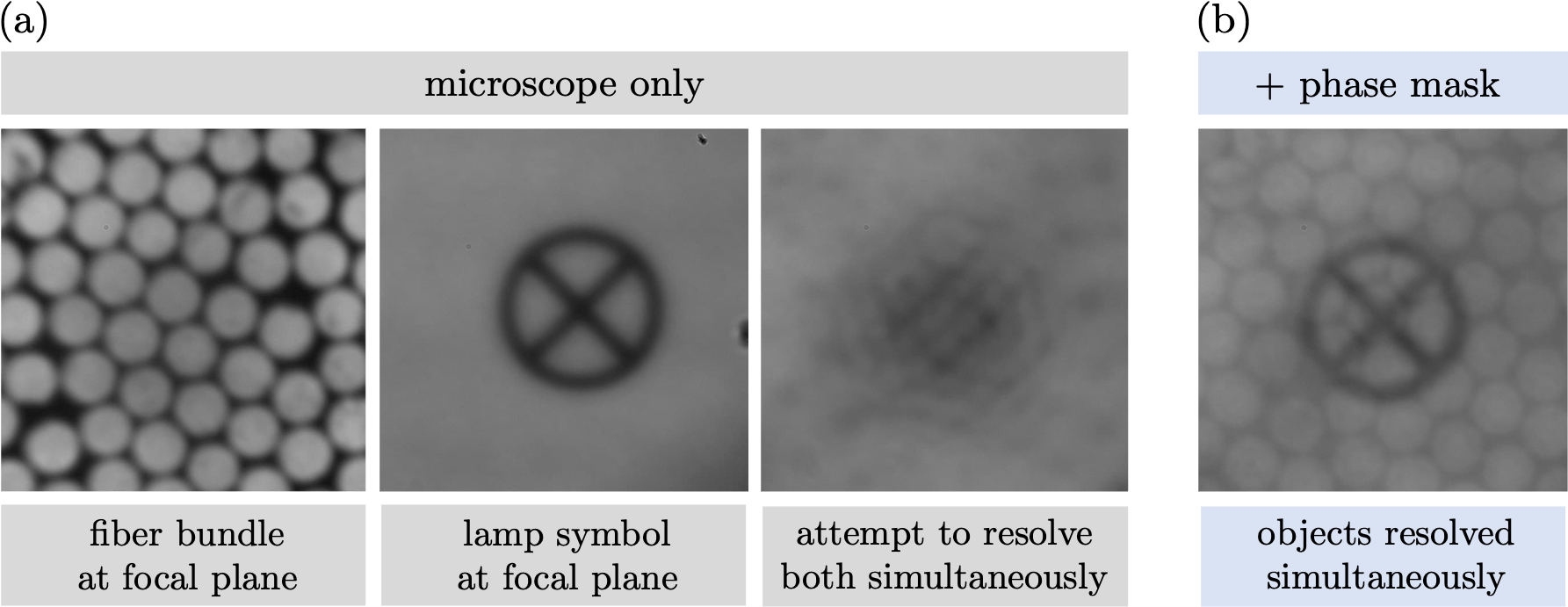}
\caption{Simultaneously imaging two objects at different distances from the focal plane of the objective. The two objects, the output facet of a fiber bundle and a lamp symbol on a transparency, are placed \qty{7.5}{\milli\meter} apart along the optical axis, and imaged simultaneously. (a) Only the microscope is used and we are unable to simultaneously resolve both objects. (b) With the inclusion of the phase mask both objects can be resolved simultaneously.} 
\label{fig:demonstration}
\end{figure}

Lastly, in Fig.~\ref{fig:demonstration}, we demonstrate how the extended depth of field enables the imaging system to simultaneously resolve objects separated by considerable distances along the optical axis, that the ordinary microscope is unable to resolve. The two objects are the output facet of a fiber bundle, and a lamp symbol, placed about \qty{7.5}{\milli\meter} apart. In addition to being an object, the fiber bundle is used to provide white-light incoherent illumination of the system. In Fig.~\ref{fig:demonstration}a (the microscope only), we see that whilst the fiber bundle and the lamp symbol can separately be brought into sharp focus, it is not possible to take an image of both objects simultaneously. Whereas when the phase mask is included (Fig.~\ref{fig:demonstration}b), these two objects are clearly discernible in the same image. This highlights the utility of the extended depth of field in simultaneous multiplane imaging. Lastly, we note that the imaging performance of the optimised device remains consistent across the visible spectrum (see Supporting Information for details).

\FloatBarrier
\section{Conclusion}

We used inverse design techniques to design a phase mask, which when placed in the Fourier plane of a simple microscope, significantly extends the depth of field without compromising the high transmission and high resolution of the microscope. The phase mask was fabricated using two-photon grayscale lithography, in combination with a pre-compensation routine, resulting in a device which matches the target design with greater than subwavelength accuracy. Through optical characterisation, we determined the enhancement factor in the depth of field to be approximately four, as quantified by the distance between the optical planes where the standard deviation of the PSF is twice its smallest value. The efficacy of the device was also demonstrated in some concrete use cases. For example, we showed how the phase mask enables simultaneous multiplane imaging, for separations between objects which were not possible with the original microscope.

We also analysed the imaging performance of the device when imaging an object at several different displacements away from the focal plane. With the phase mask inserted, we clearly saw that the key features of the objects remain discernible over larger displacements. However, this comes at the cost of reduced contrast and image brightness, artefacts such as halos, and a peak performance at the focal plane that is necessarily inferior to the ordinary microscope. The origin of these differences can be explained by examining the point spread functions. The phase mask causes an elongation in the point spread function, in addition to pronounced side lobes, and an inhomogeneous total intensity profile along the optical axis. However, we point out that contrast-enhancing and other imaging processing algorithms could be applied to minimise these artefacts. With knowledge of the PSF these could include deconvolution algorithms.

Such similarly optimised and fabricated phase masks could prove useful in existing imaging setups, where a simple method to extend the depth of field without attenuating the optical field is desired, which would naturally extend the applicability of such setups. To improve upon this work, one could invest further resources into experimenting with a wider palette of objective functions and optimisation techniques, in order to try and combat the drawbacks mentioned above. One could try relaxing the assumption of radial symmetry, and optimise a free-form structure in the $x$-$y$ plane. Alternatively, one could consider optimising a more complex optical setup; for example, by considering two or more phase masks positioned sequentially along the optical axis.

\section{Methods}

\subsection{Simulation methods}

We model the optical system using Fourier optics, which can be considered as a subset of the scalar diffraction theory of light, where the description of the propagation of light waves is based on a spatial harmonic analysis. This theory does not account for absorption or reflection, and simply describes the optical field in the transverse plane with a scalar field $E(x, y)$, which implicitly depends on the distance along the optical axis $z$ (see e.g. chapter 4 in Ref.~\citenum{SalehTeich1991}). This theory is computationally inexpensive; for example, the propagation of an initial field $\hat{E}_0(k_x, k_y)$ (in reciprocal space) to an arbitrary distance $z$ through air is given by the one-step expression $\hat{E}_z(k_x, k_y) = \hat{E}_0(k_x, k_y) \exp(-\text{i} k_z z)$, where $k_z =  \sqrt{k_0^2 - k_x^2 - k_y^2}$. We use the thin element method to approximate the optical elements, whereby the incident field $E_\text{in}$ is transformed to the outgoing field $E_\text{out}(x,y) = t_n (x, y) E_\text{in}(x, y)$, where $t(x, y)$ is the transmission coefficient through the optical element (see e.g. section 2.4 in Ref.~\citenum{SalehTeich1991} for details).

We write our simulation in Python. The simulation code is available as an installable package. We use a square simulation grid for the $x$-$y$ plane, with a width of \qty{3.2}{\milli\meter} and 1600 sample points along each axis. We implement the optical setup described in Fig.~\ref{fig:setup}, with a phase mask radius $a = \qty{1}{\milli\meter}$, and an iris-mask distance of \qty{2}{\milli\meter}. During optimisation we use the objective focal length $f_1 = \qty{20}{\milli\meter}$, compared to the \qty{45}{\milli\meter} used in the final simulations and experiments. The phase mask works just as well in both setups; we will comment more on this at the end of this section. We also introduce 10 `artificial' irises with a large radius of \qty{1.5}{\milli\meter}, equispaced between the phase mask and the objective lens. This stops the periodic boundary conditions from introducing unphysical effects, and prevents the optimiser from exploiting this. In our simulation `backwards' through the imaging system (see the subsection ``Simulations and mask design''), we do not explicitly model the propagation of a point source through the tube lens; rather, we specify an incident field on the iris as a plane wave $E(x, y) = E_0$. Our target depth of field is $L_\text{dof} = \qty{2}{\milli\meter}$. From the simulations, we generate $N_\text{slices} = 13$ slices of the field along the optical axis, centered around the object plane, at distances $z_n = -L_\text{dof} / 2 + n \delta_z$, for $n \in \{1,2,\dots,N_\text{slices}\}$, where $\delta_z = L_\text{dof} / N_\text{slices}$. Let us denote the 2D arrays corresponding to these discretised fields as $\mathbf{E}_n$. We define the objective function as follows. First we define an array that filters a narrow region around the optical axis, namely

\begin{equation}
    \mathbf{G} = \exp\left[-\left( \mathbf{X}^2 + \mathbf{Y}^2 \right) / \sigma_G^2 \right],
\end{equation}

\noindent where $\mathbf{X}$ and $\mathbf{Y}$ are the two-dimensional NumPy mesh grids corresponding to the discretised $x$ and $y$ coordinates, and we choose $\sigma_G = \qty{5}{\micro\meter}$. The filtered and summed intensity of a single field array is then $I_n = \text{sum}\left( \text{abs}(\mathbf{E}_n)^2 \right)$. From this, we define the objective function as

\begin{equation}
    \text{objective} = \sum_{n=0}^{N_\text{slices} - 1} I_n - (W / \delta_z) \sum_{n=1}^{N_\text{slices} - 1} \text{abs}\left( I_n - I_{n-1} \right),
\end{equation}

\noindent where we choose the weighting term $W = 330$.

For the Bayesian optimisation, we use the software package \code{bayesian-optimization} \cite{bayesian-optimization}, available \href{https://github.com/bayesian-optimization/BayesianOptimization}{here}. Bayesian optimisation is a global optimisation technique that approximates the landscape of the true objective function to find the global optimum. It strategically balances exploration of the search space and exploitation of found local maxima, to efficiently converge on the best solution. With this optimisation technique, we parameterise the radial profile of the design with a 6th order Chebyshev polynomial, and optimise the coefficients of the polynomial. We seed the process with 1000 initial points and then proceed for a further 100 iterations. Already this is enough to find a good starting point for the next step of the hybrid optimisation approach. 

For the gradient descent based local optimisation, we use the software package \code{nlopt} \cite{JohnsonNLopt}, available \href{https://github.com/stevengj/nlopt}{here}. Specifically, we use the method \code{NLOPT\_LD\_MMA}, which is an implementation of globally convergent method-of-moving-asymptotes \cite{Svanberg2002}. To calculate the gradients of the fields resulting from our simulation with respect to the design variables that parameterise the radial height profile of the phase mask, we use the software package \code{jax} \cite{jax2018github}, available \href{https://github.com/google/jax}{here}. JAX can automatically differentiate native Python and NumPy functions. Therefore, if a simulation only uses native Python and (one of the many supported) NumPy functions, one just needs to swap \code{numpy} functions with \code{jax.numpy} functions, and then calculate the gradient of an arbitrary (functionally pure) function with the \code{jax.grad} utility. We also exploited the fact that JAX supports just-in-time compilation and execution on GPUs.

Our three-step hybrid optimisation method works as follows. First, we use Bayesian optimisation (with Chebyshev polynomial parameterisation) as described above, and pick the 20 best designs (as quantified by the figure of merit). For each of these 20 starting designs, we perform a further two steps and choose the best resulting design out of the 20 final designs. In the second step, we also parameterise the radial height profile of the design with a Chebyshev polynomial; however, we use a 10th order polynomial and perform gradient-based optimisation. In the third step, we switch to a radial Cartesian grid with 50 sample points and optimise the radial height profile directly with gradient-based optimisation.

As mentioned above, we optimise the phase mask using an objective focal length $f_1 = \qty{20}{\milli\meter}$, instead of \qty{45}{\milli\meter} that is used in the experiments. However, the calculations of the field distributions shown in Fig.~\ref{fig:simulation} are performed assuming a \qty{45}{\milli\meter} objective focal length. The phase mask still works just as well, but the numerical aperture of the system is smaller, and consequently, the target depth of field is increased from $L_\text{dof} = \qty{2}{\milli\meter}$ during optimisation, to $L_\text{dof} = \qty{10}{\milli\meter}$ during the final simulations shown in Fig.~\ref{fig:simulation}. The actual depth of field that we achieve in the experiments is \qty{11.8}{\milli\meter}. We experimented with longer target depth of fields, but found that this led to a PSF that was increasingly inhomogeneous along the optical axis. However, it is possible that these problems could be minimised by further tuning of the figure of merit.

\subsection{Fabrication methods}

The phase mask was fabricated using two-photon grayscale lithography in a Quantum X machine (Nanoscribe GmbH \& Co. KG). To obtain a low surface roughness, we printed the phase mask using Nanoscribe's IP-S photoresist and a 25x/NA0.8 microscope objective lens (Zeiss, LD LCI Plan-Apochromat Imm Corr DIC M27). For convenience during sample handling, the phase masks were printed on adhesive microscopy slides (ThermoScientific, Menzel-Gläser, SuperFrost Plus). Due to the limited size of the field of view in the printing setup, the phase mask has to be split into different print fields with a size of \qtyproduct{420 x 420}{\micro\metre}. These sections are printed sequentially and are stitched together using Nanoscribe's 2GL stitching routine. After printing, the samples are developed in propylene glycol methyl ether acetate (PGMEA) for twenty minutes and twice in isopropanol for five minutes each.

When naively printing the phase mask with the calculated height profile as the target structure, as depicted in Fig.~\ref{fig:fabrication}a, the resulting structures show systematic imperfections like stitching seams, optical tilt in each print field, as well as global and local shrinkage. In addition, the quality of the fabricated structures is reduced by statistical imperfections such as axial print field offsets due to a limited precision of the automated interface finder, inhomogeneities in the photoresist (and therefore also in the cross-linked polymer), and micro-explosions due to locally generated heat or dust particles. These systematic and statistical imperfections are reduced with a pre-compensation routine as described in the main text.

A reflection-based spinning-disk confocal microscope (MarSurf CM explorer S/W, Mahr GmbH) was used to measure height profiles, due to its high lateral and axial resolution, as well as the non-invasive nature of this method. Nonetheless, it is also important to note that the measurement method itself introduces additional measurement artefacts to the result. They include deviating axial and lateral offsets for different measurement fields, as well as substrate rotation and tilt. To support the automated lateral alignment routine of the microscope software, small aperiodic alignment frames were printed around each structure.

For calculating the difference map, it is crucial that the measured height profile and the design structure are well aligned. Therefore, the measured height profile is tilt- and offset-corrected by fitting a plane through the measurement points on the substrate. Afterwards, the alignment of the measured height profile and target design is optimised regarding lateral displacements, lateral scaling, and axial rotation using image cross-correlation techniques \cite{pre-compensation}. In principle, multiple iteration steps of this pre-compensation routine can still be performed, yielding slightly different print results. It was observed, however, that the resulting height profiles from additional iteration steps tend to oscillate around the ideal target design with amplitudes smaller than \qty{100}{\nano\meter}.

\begin{acknowledgement}

T.J.S. acknowledges funding from the Alexander von Humboldt Foundation.
M.N. and C.R. acknowledge support by the KIT through the ``Virtual Materials Design'' (VIRTMAT) project.
K.P., P.H. and A.S. acknowledge funding from the Research Council of Finland Flagship Programme, Photonics Research and Innovation (PREIN), decision number 346529.
H.P. acknowledges funding from the Finnish Foundation for Technology Promotion (Grant No. 9085).
S.K., C.R. and M.W. acknowledge funding by the Deutsche Forschungsgemeinschaft (DFG, German Research Foundation) under Germany’s Excellence Strategy for the Excellence Cluster “3D Matter Made to Order” (2082/1–390761711), by the Carl Zeiss Foundation, and by the Helmholtz program Materials Systems Engineering. Parts of the work were performed on the HoreKa supercomputer funded by the Ministry of Science, Research and the Arts Baden-Württemberg and by the Federal Ministry of Education and Research.

\end{acknowledgement}

\begin{suppinfo}

Supporting information available:

\begin{enumerate}
    \item An analysis of the wavelength dependence of the imaging system
    \item A description of how the simulated images are formed, and an explanation of optical reciprocity.
    \item Simulations that use microscopy measurements of the height profile of the fabricated device.
\end{enumerate}

\noindent This material is available free of charge via the internet at \href{http://pubs.acs.org}{http://pubs.acs.org}

\end{suppinfo}

\bibliography{references}

\end{document}


\maketitle

\section{Wavelength dependence of the imaging setup}

\begin{figure}[h]
    \centering
    \includegraphics[width=0.7\linewidth]{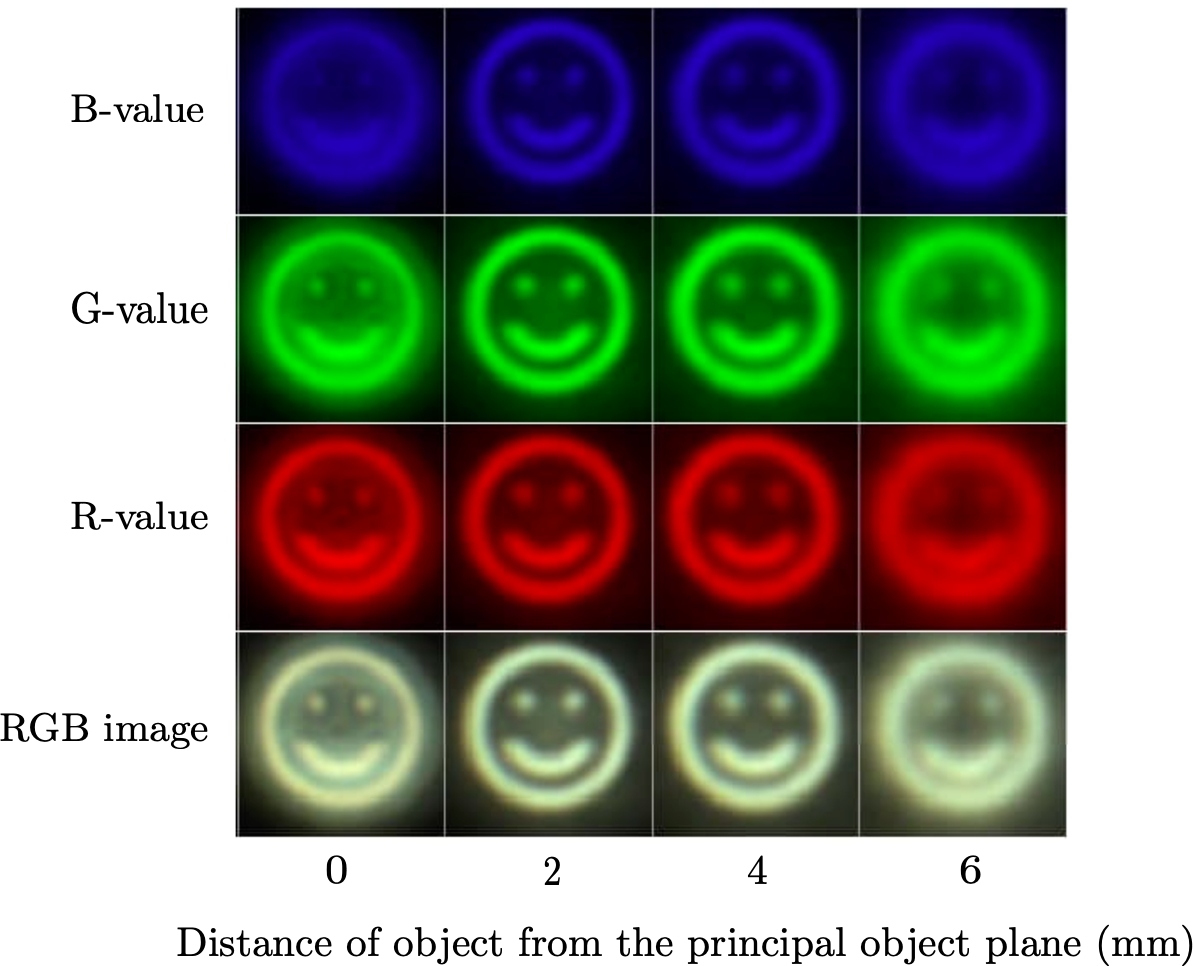}
    \caption{Images of transparencies as they are moved out of focus, along the $z$-axis, under white light incoherent illumination. The top three rows show the blue, green, and red channels, whereas the bottom row shows the full colour image. We see that the imaging system is rather wavelength insensitive, with no particularly obvious differences between the colour channels.}
    \label{fig:smiles}
\end{figure}

In Fig.~\ref{fig:smiles} we visually analyse the wavelength dependence of the imaging system with the phase mask inserted. The images shown are a sample from Fig.~5 in the main text. However, in the main text we only show grayscale images corresponding to the red-channel of RGB images, whereas here we show all channels (top 3 rows), as well as the full colour image (bottom row). We see that the imaging performance is rather wavelength insensitive, meaning that the phase mask is broadband, covering the visible spectral range and suitable for white light illumination widely used in microscopes.

\section{Image formation and reciprocity}
\label{sec:image-formation}

The image formed by the setup in the image plane (where the camera is) can be calculated as follows. First, consider a single point source on the image plane at $(x_i,y_i)$. The intensity generated on or near the object plane is the PSF that we also show in Figs.~2 and 4 in the main text; let us call it $P(x,y,z,x_i,y_i)$. Note that the system is not shift-invariant; the PSF in general depends on the distance from the optical axis. Optical reciprocity now dictates that if we place a point source at $(x,y,z)$ near the object plane, its intensity at the point $(x_i,y_i)$ on the image plane should be equal to $P(x,y,z,x_i,y_i)$.

Therefore, the PSF tells us how much intensity the points $(x,y,z)$ on the object side of the setup contribute to the image at the point $(x_i,y_i)$ on the image plane. The image intensity is then (assuming incoherent imaging)
\begin{equation}
    I_\text{im}(x_i,y_i) = \int I_\text{ob}(x,y,z) P(x,y,z,x_i,y_i) \mathrm{d}x \mathrm{d}y \mathrm{d}z ,
\end{equation}
where $I_\text{ob}(x,y,z)$ is the intensity distribution of the object. This equation can be used to generate simulated images, but more importantly, it underlines why we chose to work with the object-side PSF with image-plane point sources rather than the other way around.

\section{Simulations with the fabricated devices}

\begin{figure}[h]
    \centering
    \includegraphics[]{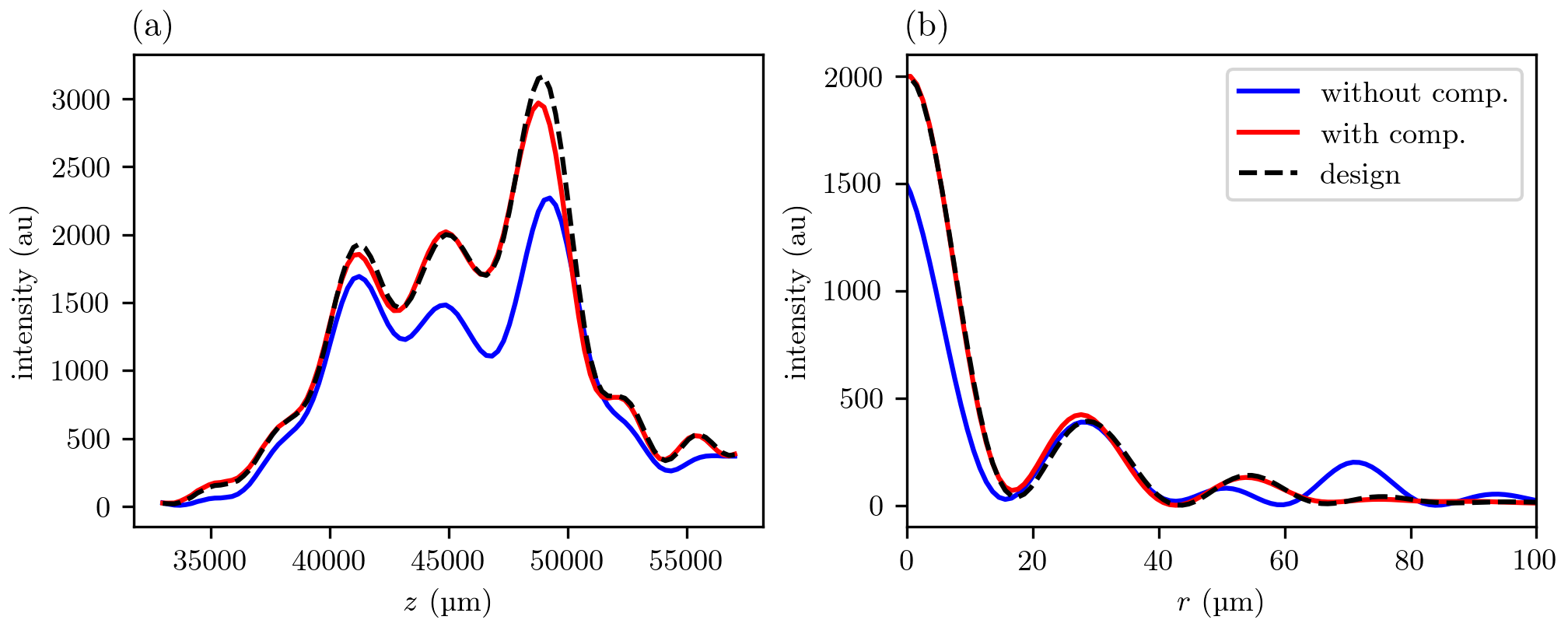}
    \caption{Simulations of the fabricated device height profile, both with (red) and without (blue) the pre-compensation routine, as compared to the original design (black). Cuts of the intensity distribution (a) along the optical axis at $r=0$, and (c) along the radial axis at $z=0$.}
    \label{fig:comparison}
\end{figure}

To determine whether we can expect the fabricated device to perform as expected, we perform simulations with the height profile as determined by microscopy measurements. Slices of the intensity profile are shown in Fig.~\ref{fig:comparison}, both with (red) and without (blue) the pre-compensation routine, as compared to the original design (black). We note that the pre-compensation routine was indeed necessary to match the PSF of the original design. We see a strong quantitative agreement.

\section{Full-width half-maximum of the PSF}

\begin{figure}[t!]
    \vspace*{-6in}
    \centering
    \includegraphics[]{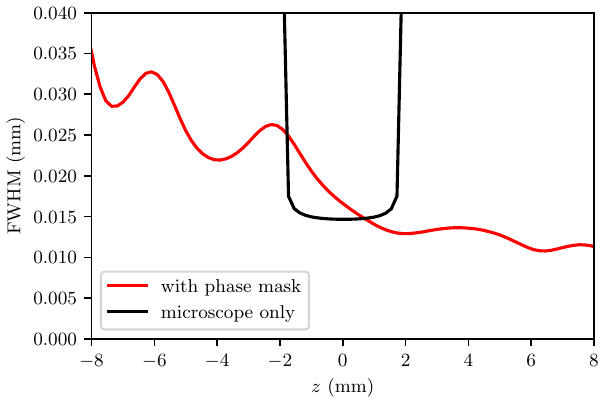}
    \caption{Variation in the full-width half-maximum (FWHM) of the central peak of the PSF along the optical axis, for the system with (red line) and without (black line) the optimised phase mask.}
    \label{fig:FWHM}
\end{figure}

In Fig.~\ref{fig:FWHM} we see a comparison between the full-width half-maximum (FWHM) of the microscope with and without the optimised phase mask inserted. At the focal point, the FWHM is only slightly increased by the presence of the phase mask. In this sense, the resolution is only minimally compromised. Note that the FWHM of the plain microscope diverges at approximately \qty{2}{\milli\meter} away from the focal point. This is because the central peak collapses and the FWHM is no longer meaningful. Whereas, with the phase mask inserted, due to the extended depth of field, the central peak only collapses at much larger distances. It is also interesting to note the asymmetric and non-monotonic variation in the FWHM about the focus point. This fits in with the inhomogeneous imaging performance discussed in the main text.

